\documentstyle[12pt]{article}
\begin{document}

\thispagestyle{empty}
\begin{flushright}
JHU-TIPAC-95010\\
BONN-TH-95-08\\
hepth@xxx/9504124\\
April 1995
\end{flushright}

\bigskip\bigskip\begin{center} {\bf
\Large{Supersymmetric sigma models and the 't Hooft instantons}}
\end{center} \vskip 1.0truecm

\centerline{\bf A. Galperin}
\vskip5mm
\centerline{Department of Physics and Astronomy}
\centerline{Johns Hopkins University}
\centerline{Baltimore, MD 21218, USA}
\vskip5mm
\centerline{\bf and}
\vskip5mm
\centerline{\bf E. Sokatchev}
\vskip5mm
\centerline{Physikalishes Institut}
\centerline{Universit\"at Bonn}
\centerline{Nu{\ss}allee 12, 53115 Bonn, Germany}
\vskip5mm

\bigskip \nopagebreak \begin{abstract}
\noindent
Witten's linear sigma model for ADHM instantons possesses a natural
$(0,4)$ supersymmetry. We study generalizations of the infrared limit of the
model that are invariant under  $(4,4)$
supersymmetry. In the case of four space-time dimensions a background with
a conformally flat metric and torsion is required. The geometry is
specified by a single
real scalar function satisfying Laplace's equation.
It gives rise to 't Hooft instantons for the gauge group $SU(2)$, instead
of the general ADHM instantons for an $SO(n)$ gauge group in the case
$(0,4)$.
\end{abstract}

\newpage\setcounter{page}1

\section{Introduction}

Recently, Witten \cite{Witten} constructed a $(0,4)$ supersymmetric linear
sigma model in two dimensions with a potential term. Remarkably, under
simple assumptions about the structure of the $(0,4)$ multiplets involved,
the Yukawa couplings of the model satisfy the ADHM equations of the
instantons construction \cite{Adhm}.  In the infrared limit the massless
chiral fermions in the model are coupled to an instanton gauge field,
obtained according to the ADHM prescription \cite{Adhm}.  The fact that $(0,4)$
supersymmetry is related to self-duality of the target-space gauge fields
is not new (see, for instance,
\cite{Howe}). However, it is a very unusual feature of Witten's model
that it restricts the general self-dual gauge fields to those with finite
action, i.e., {\it instantons}. In \cite{GalSok} we gave a manifestly
supersymmetric
and off-shell formulation of the ADHM sigma model in harmonic $(0,4)$
superspace.

In the present paper we study generalizations of Witten's model to $(4,4)$
supersymmetry. More precisely, we examine the conditions for  the model to have
off-shell $(4,4)$ supersymmetry in the infrared limit.  A necessary condition
is that
the number of  massless chiral fermions  be tied to the dimension of the target
space-time. In this case the linearized action of the massless sector
possesses an off-shell $(4,4)$ supersymmetry. We find the most general
action of the massless fields consistent with this supersymmetry. It
necessarily
involves nontrivial space-time geometry with torsion. In four space-time
dimensions
 supersymmetry restricts the Yang-Mills gauge group to $SU(2)$ and the
metric to be conformally flat.  The conformal factor satisfies
Laplace's equation. In fact, it is identified with the scalar field of the
't Hooft construction of the $5k'$ family of $SU(2)$ instantons \cite{Hooft}.
We also find
analogs of the 't Hooft instantons in higher dimensions.

In section \ref{sec2} we briefly review the harmonic superspace
construction of the ADHM sigma model \cite{GalSok}.
In section \ref{sec3} we concentrate on its infrared limit and generalize it to
include a
nontrivial space-time background. Then we study the conditions for the massless
theory
to possess off-shell $(4,4)$ supersymmetry.  In section 4 we derive the
component action and discuss further problems. In the concluding section we
compare
our results with various facts  about (4,4) sigma models known in the
literature.

\section{ADHM sigma model in $(0,4)$ harmonic superspace}\label{sec2}

In this section we shall review the construction of the ADHM sigma model
with manifest $(0,4)$ supersymmetry. More details can be found in
\cite{GalSok}. The supersymmetry algebra is given by
\begin{equation}\label{(0,4)algebra}
\{Q_-^{AA'}, Q_-^{BB'}\}=2\epsilon^{AB}\epsilon^{A'B'} P_{--}
\end{equation}
 Here $\pm$ indicate the Lorentz
($SO(1,1)$) weights, and we always write the weights as lower indices. $A$
and $A'$ are doublet indices of the $SO(4)\sim SU(2)\times SU(2)'$
automorphism group of (\ref{(0,4)algebra}). The supercharges
 satisfy the reality condition $\overline{Q_-^{AA'}} =
\epsilon_{AB}\epsilon_{A'B'} Q_-^{BB'}$. The algebra (\ref{(0,4)algebra})
can be realized on the super-world sheet with coordinates $x_{++}, x_{--},
\theta^{AA'}_+$:
\begin{equation}\label{(0,4)SUSY}
\delta x_{++}=-i\epsilon^{AA'}_+\theta_{+AA'}, \;\;
\delta x_{--}=0, \;\; \delta\theta^{AA'}_+=\epsilon^{AA'}_+.
\end{equation}

Following the method of $SU(2)$ harmonic superspace \cite{harms}, we
extend the world sheet by means of the harmonic variables $u^{\pm A} \in
SU(2)/U(1): \ \ u^{+A}u^-_A=1, \ \ u^-_A= \overline {u^{+A}}$.  These
variables are inert under supersymmetry, transform as doublets of one of the
$SU(2)$
subgroups of automorphisms, and are defined modulo the $U(1)$ subgroup of
$SU(2)$. In essence, harmonics parametrize the two-sphere $S^2$ of all
possible choices of the three complex structures characteristic for $(0,4)$
supersymmetry. The extended superspace $x_{++}, x_{--},
\theta^{AA'}_+, u^{\pm A}$ possesses an invariant subspace of lower
Grassmann dimension:
\begin{eqnarray}
\hat x_{++} = x_{++} + i\theta^{AA'}_+\theta^{B}_{+A'} u^+_{(A}u^-_{B)},
\ \ x_{--}, \ \ \theta^{+ A'}_{+}= u^+_A \theta^{AA'}_+, \ \ u_A^\pm\; ;
\label{bas} \\
\delta \hat x_{++}=-2i\epsilon^{AA'}_+u^-_A\theta^+_{+A'}, \;\;
\delta x_{--}=0, \;\; \delta\theta^{+A'}_+=\epsilon^{AA'}_+u^+_A\; .
\end{eqnarray}
Note that in the notation $\theta^{\pm }_{+}$ the upper index $\pm$ is a $U(1)$
and the
lower one is Lorentzian. It is this {\it analytic} superspace that is most
adequate for
describing multiplets of $(0,4)$ supersymmetry. The analytic superfields
$\Phi^q
(x,\theta^+,u)$ \footnote{To simplify the notation we shall not write
explicitly $\hat x_{++}$ when it is clear that we work in the analytic
superspace (\ref{bas}).} carry in general a  $U(1)$ harmonic charge and an
$SO(1,1)$ weight.  They have a very
short Grassmann expansion, e.g.,
\begin{equation}\label{expa}
\Phi^q(x,\theta^+,u) = \phi^q(x,u) + \theta^{+A'}_+\xi^{q-1}_{-A'}(x,u)
+(\theta^+_+)^2 f^{q-2}_{--}(x,u)\; ,
\end{equation}
where $(\theta^+_+)^2 \equiv \theta^{+A'}_+\theta^+_{+A'}\;$.  The
coefficients in (\ref{expa}) are harmonic-dependent fields; they can be
expanded in terms of spherical harmonics on $S^2$.
For certain values of the $U(1)$ charge $q$ such superfields can be made real
in the
sense of a special conjugation on $S^2$ (for more details about the harmonic
formalism see \cite{harms}, \cite{GalSok}).

Finally, to complete the formalism we need the expression for the covariant
harmonic  derivative in the analytic
superspace (\ref{bas}):
\begin{equation}\label{anD++}
D^{++} = u^{+A}{\partial\over\partial u^{-A}} + i(\theta^+_+)^2
{\partial\over\partial\hat x_{++}}\; .
\end{equation}
It will help us to write down irreducibility conditions and equations of motion
for
harmonic superfields.

The ADHM sigma model of Witten exploits three types of $(0,4)$ multiplets:
chiral (right-handed) fermions and two
different scalar multiplets that
contain both bosons and left-handed fermions.  The chiral fermions are
described in harmonic superspace by the following {\it anticommuting} and
real (in the sense mentioned above) superfields
\begin{equation}\label{Lam}
\Lambda^a_+(x,\theta^+,u) = \lambda^a_+(x,u) + \theta^{+A'}_+s^{-a}_{A'}(x,u)
+i(\theta^+_+)^2  \sigma^{--a}_{-}(x,u)\;
\end{equation}
with the free action
\begin{equation}\label{acL}
S_\Lambda = {1\over 2}\int d^2x du d^2\theta^+_+ \;
\Lambda^{a}_+D^{++} \Lambda^a_{+}\; .
\end{equation}
The external index $a=1,\ldots,n+4k'$ is an $SO(n+4k')$ one. The free
equation of motion $D^{++} \Lambda^a_{+}=0$ shows that all the components
of $ \Lambda^a_{+}$ are auxiliary, except for the lowest component in the
harmonic expansion of $\lambda^a_+(x,u)$, i.e.,
$\lambda^a_+(x,u)=\lambda^a_+(x)$.  These are the physical chiral fermions
satisfying the free equation $\partial_{--}\lambda^a_+(x)=0$ and forming a
trivial on-shell representation of $(0,4)$ supersymmetry.
However, as shown in \cite{GalSok}, its off-shell
version requires the infinite sets of auxiliary fields contained
in (\ref{Lam}), except
if the number of chiral fermions is a multiple of four (these ``short''
multiplets were discussed in the second ref. \cite{Howe}, see also
\cite{Gates}).

One of the scalar multiplets (we shall call it non-twisted) involves the
coordinates $X^{AY}$ of the Euclidean target space $R^{4}$,  in which the
Yang-Mills fields will be defined.  \footnote{We shall discuss the
generalization to the case
of $R^{4k}$ at the end of section 3.} It is described by the real analytic
superfield
$X^{+Y}(x,\theta^+,u)
\;\;(Y=1,2 $ is an $Sp(1)\sim SU(2)$ index), which satisfies the following
harmonic
irreducibility condition
\begin{equation}\label{irr}
D^{++} X^{+Y} = 0\; .
\end{equation}
The solution to it
\begin{equation}\label{X}
X^{+Y}(x,\theta^+,u) = X^{AY}(x)u^+_A + i\theta^{+A'}_+\psi^Y_{-A'}(x)
-i(\theta^+_+)^2 \partial_{--}X^{AY}(x)u^-_A
\end{equation}
contains $4$ bosonic and $4$ fermionic real off-shell fields.  The free action
for the
this multiplet is again given as an integral over the analytic superspace:
\begin{equation}\label{acX}
S_X = i\int d^2x du d^2\theta^+_+ \; X^{+Y}\partial_{++} X^+_Y\; .
\end{equation}

Finally, the last multiplet is the so-called ``twisted" scalar multiplet, in
which the $SU(2)$ indices carried by the bosons and fermions are
interchanged (as compared to the non-twisted multiplet). Its
superspace description requires a set of {\it anticommuting abelian gauge}
superfields $\Phi^{+Y'}_+(x,\theta^+,u)$  $(Y'=1,2,\ldots, 2k'$ is an $Sp(k')$
index).
The gauge transformations have the form
\begin{equation}\label{gau}
\delta \Phi^{+Y'}_+ = D^{++} \omega^{-Y'}_+
\end{equation}
with analytic parameters $\omega^{-Y'}_+(x,\theta^+,u)$. Using these
parameters one can choose the ``harmonically short'' and
non-manifestly-supersymmetric  {\it
Wess-Zumino-type gauge}
\begin{equation}\label{WZ}
\Phi^{+Y'}_+(x,\theta^+,u) = \theta^{+A'}_+\phi^{Y'}_{A'}(x)
+i(\theta^+_+)^2  u^{-}_A \chi^{Y'A}_{-}(x)\; .
\end{equation}
in which only the $4k'$ real physical bosons $\phi^{Y'}_{A'}$ and fermions
$\chi^{Y'A}_{-}$ are left. This multiplet is off shell.  The gauge invariant
free action for
the  superfield $\Phi^+_+$ has been given in \cite{GalSok} and we shall not
need it here.

Now we turn to the discussion of the potential-type coupling of the above three
$(0,4)$ multiplets. Such a coupling is severely restricted by dimension,
 $SO(1,1)$ and  $U(1)$ invariance, as well as Grassmann analyticity. An
important
additional assumption made by Witten in \cite{Witten} is that the part $SU(2)'$
of the
$(0,4)$ supersymmetry automorphism group is preserved (this requirement is
motivated by the desire to obtain a CFT in the infrared limit). As shown in
\cite{GalSok}, then the only possible coupling term is
\begin{equation}\label{int}
S_{int} = m \int d^2x du d^2\theta^+_+ \;  \Phi^{+Y'}_+ v^{+a}_{Y'}(X^+,u)
\Lambda_+^a\; .
\end{equation}
It is invariant under the gauge transformation (\ref{gau}) (together with the
kinetic term (\ref{acL}) for $\Lambda^a_+$) provided the chiral fermions
transform as
\begin{equation}\label{gauLv}
\delta \Lambda^a_{+} = mv^{+a}_{Y'}(X^+,u)  \omega^{-Y'}_+\; ,
\end{equation}
and the matrix $v^{+a}_{Y'}(X^+,u)$ satisfies the following two conditions
\begin{equation}\label{c1}
v^{+a}_{Y'}v^{+a}_{Z'} = 0\; ,
\end{equation}
\begin{equation}\label{c2}
D^{++} v^{+a}_{Y'}(X^+,u) = 0\; .
\end{equation}

The general solution to (\ref{c2}) is  (recall (\ref{irr}))
\begin{equation}\label{lin}
v^{+a}_{Y'}(X^+,u) = u^{+A} \alpha^a_{AY'} + \beta^a_{Y'Y} X^{+Y}\; ,
\end{equation}
where the matrices $\alpha,\beta$ are constant. At $\theta^+=0$, the
matrix $v^{+a}_{Y'}(X^+,u)$ reduces to
\begin{equation}\label{lin'}
v^{+a}_{Y'}(X^+,u)|_{\theta=0} = u^{+A} (\alpha^a_{AY'}  +
\beta^a_{Y'Y} X^{Y}_A) \equiv u^{+A}
\Delta^a_{AY'}\; ,
\end{equation}
and then the  other condition (\ref{c1}) implies for the matrix
$\Delta^a_{AY'}$
\begin{equation}\label{ADHM}
\Delta^a_{AY'}\Delta^a_{BZ'} + (A\leftrightarrow B)= 0\; .
\end{equation}
The matrix $\Delta^a_{AY'}$, linear in $X$ and satisfying (\ref{ADHM})
is the starting point in the ADHM construction for instantons \cite{Adhm}.

Now we discuss the infrared limit of the theory. To this end one has to
separate the massless and massive modes. Among the $n+4k'$ left-handed
fermions $\lambda^a_+$ contained in $\Lambda^a_+$ there is a subset of $4k'$
which are paired with the right-handed fermions in $\Phi$ and become
massive (together with the bosons from $\Phi$). The remaining chiral
fermions stay massless. To diagonalize the action, we complete the
$2k'\times (n+4k')$ matrix $v^{+a}_{Y'}(X^+,u)$ to a full {\it orthogonal}
matrix $v^{\hat aa}(X^+,u)$, where the $n+4k'$ dimensional index $\hat a =
(+Y', -Y',i)$ and $i=1,\ldots, n$ is a vector index of the group $SO(n)$.
Orthogonality means
\begin{equation}\label{or}
v^{\hat aa} v^{\hat b a} = \delta^{\hat a\hat b}\; ,
\end{equation}
where $\delta^{+Y', -Z'} = - \delta^{-Y', +Z'} = \epsilon^{Y'Z'}, \;\;
\delta^{+Y', +Z'}=\delta^{-Y', -Z'}=\delta^{\pm Y', i}=0 $.
Since $v^{+a}_{Y'}$ is a function of $X^{+Y}$ and
$u^\pm$, we take the other blocks of $v^{\hat aa}$, namely $v^{-Y'a}$
and $v^{ia}$ to be such functions too.

With the help of the matrix $v^{\hat aa}$ we can make a change of variables
from the superfield $\Lambda^a_+$ to $\Lambda^{\hat a}_+ =v^{\hat
aa}\Lambda^a_+$. Then the gauge transformation (\ref{gauLv}) gets the form
\begin{equation}\label{gauL-}
\delta\Lambda^{-Y'}_+ = m\omega^{-Y'}_+, \ \ \ \delta\Lambda^{+Y'}_+ =
\delta\Lambda^{i}_+ = 0\; ,
\end{equation}
hence the superfields $\Lambda^{-Y'}_+$ can be completely gauged away.
Further, the superfields $\Lambda^{+Y'}_+$ enter the action without
derivatives; their elimination results in the following Lagrangian
for the chiral fermions
\begin{eqnarray}
{\cal L}^{++}_{++}(\Lambda)
&=& {1\over 2} \Lambda^i_+[\delta^{ij} D^{++} + (V^{++})^{ij}]  \Lambda^j_+
\nonumber \\
&-& \label{LL}
 {1\over 2}[(V^{++})^{-i}_{Y'} \Lambda^{i}_+ + m\Phi^+_{+Y'}] (V^{-1})^{Y'Z'}
[(V^{++})^{-j}_{Z'} \Lambda^{j}_+ + m\Phi^+_{+Z'}]\; .
\end{eqnarray}
Here we used the notation
\begin{equation}\label{VV}
(V^{++})^{\hat a\hat b} = v^{\hat aa}D^{++} v^{\hat ba}\; , \ \ \
V_{Y'Z'} =(V^{++})^{--}_{Y'Z'}\; .
\end{equation}

In the infrared limit  $m\rightarrow\infty$ the kinetic term for $\Phi^+_+$ is
suppressed, so the second line of  (\ref{LL}) becomes auxiliary and can be
dropped. The
final result for the massless sigma model is:
\begin{equation}\label{ADga}
S_{m\rightarrow\infty} = \int d^2x du d^2\theta^+_+ \; [iX^{+Y}\partial_{++}
X^+_Y + iP^{-Y}_{++} D^{++} X^+_Y  + {1\over 2} \Lambda^i_+(\delta^{ij} D^{++}
+
({V}^{++})^{ij})  \Lambda^j_+]\; .
\end{equation}
Here we have added the kinetic term (\ref{acX}) for $X^{+Y}$ and have
introduced the
harmonic irreducibility condition (\ref{irr}) into the action with the Lagrange
multiplier
$P^{-Y}_{++}$. The object
\begin{equation}\label{calV} ({ V}^{++})^{ij}(X^+,u) =
v^{ia}(X^+,u)D^{++}v^{ja}(X^+,u)\; .
\end{equation}
is the twistor transform of the ADHM $SO(n)$ gauge field (or, we should
rather say, the harmonic version \cite{harW} of Ward's \cite{Ward}
instanton construction).

The alternative to the manifestly supersymmetric gauge $\Lambda^{-Y'}_+=0$
above is the
Wess-Zumino gauge (\ref{WZ}). In it, after a suitable diagonalization and again
in the
infrared limit one finds an ADHM gauge field coupled to the massless subset of
the chiral
fermions $\lambda^a_+$ \cite{Witten}.

This completes our review of the superfield construction for
the ADHM sigma model.

\section{Searching for (4,4) supersymmetry}\label{sec3}

The procedure of the  previous section lead to the massless action (\ref{ADga})
for the
superfields
$X^{+Y}$ and $\Lambda^i_+$. It involves four real bosons $X^{AY}$
and four real left-handed  fermions $\psi^{A'Y}_-$ coming from the superfield
$X^{+Y}$,
as well as
$n$ real right-handed fermions $\lambda^i_+$ from the matter superfields
$\Lambda_+^i$. If we want to form a (4,4) multiplet out of them, the first
necessary condition is to match the numbers of left- and right-handed
fermions. Consequently, we have to choose $n=4$ and restrict the gauge group
to (at most) $SO(4)\sim SU(2)\times SU(2)$.

The search for further (4,4) supersymmetry is based on an examination of the
flat action obtained by putting $V^{++}=0$ in (\ref{ADga}):
\begin{equation}\label{freeact}
S_{\mbox{free}} = \int d^2xdud^2\theta^+_+\;  \left(i X^{+Y}\partial_{++} X^+_Y
+i P^{-Y}_{++}
D^{++} X^+_Y  + {1\over 2} \Lambda^{i}_+ D^{++}\Lambda^i_{+} \right) \; .
\end{equation}
It is not hard to check that this free action has two different off-shell
(4,4) supersymmetries,
depending on how the $SU(2)$ indices are involved in the transformation laws.
The first possibility is obtained by replacing the $SO(4)$ vector index $i$ by
the
$SU(2)\times SU(2)$ pair $\dot AY$ ($\dot A$ is an $SU(2)$ index of a new
type):
\begin{eqnarray}
\delta X^{+Y} &=& i\varepsilon^+_{-\dot A} \Lambda^{\dot AY}_+ \; , \nonumber
\\
\delta \Lambda^{\dot AY}_+ &=& -2\varepsilon^{-\dot A}_- \partial_{++} X^{+Y} -
\varepsilon^{+\dot A}_-
 P^{-Y}_{++} \; ,  \nonumber \\
 \delta P^{-Y}_{++} &=& -2i\varepsilon^-_{-\dot A}  \partial_{++} \Lambda^{\dot
AY}_+  \; ,
 \label{freetr}
 \end{eqnarray}
where $\varepsilon^{\pm \dot A}_- = u^\pm_A \varepsilon^{A\dot A}_-$. Actually,
these
transformation laws originate from the $\theta^+_-$ expansion of the (4,4)
superfield
$q^{+Y} (\theta^+_+, \theta^+_-)$ obtained by dimensional reduction from the
$N=2\;
D=4$ hypermultiplet \cite{harms}:
\begin{equation}
q^{+Y} = X^{+Y}  + i\theta^+_{-\dot A} \Lambda_+^{\dot AY}  -{i\over 2}
(\theta^+_-)^2 P^{-Y}_{++} \; .
\end{equation}
Moreover, in this case the free action (\ref{freeact}) itself can be derived
from the
hypermultiplet action
\begin{equation}
S = \int d^4xdud^4\theta^+\;  q^{+Y}D^{++}q^+_Y \; .
\end{equation}

 The second  possibility is obtained by writing $i$ as $A\dot Y$ (now $\dot Y$
is another
type of $SU(2)$ index) and then decomposing
$\Lambda^{A\dot Y}_+$ into harmonic $U(1)$ projections $\Lambda^{\pm\dot Y}_+ =
u^\pm_A
\Lambda^{A\dot Y}_+$ ($\Lambda_+^{\pm\dot Y}$ should not be confused with
$\Lambda_+^{\pm Y'}$ from section 2):
\begin{eqnarray}
\delta X^{+Y} &=& i\varepsilon^{Y\dot Y}_{-} \Lambda^{+}_{+\dot Y} \; ,
\nonumber \\
\delta \Lambda^{+\dot Y}_+ &=& -2\varepsilon^{Y\dot Y}_- \partial_{++} X^{+}_Y
\; ,
\nonumber \\
\delta \Lambda^{-\dot Y}_+ &=& \varepsilon^{Y\dot Y}_-  P^{-}_{++Y} \; ,
\nonumber \\
 \delta P^{-Y}_{++} &=& -2i\varepsilon^{Y\dot Y}_{-}  \partial_{++}
\Lambda^{-}_{+\dot Y}
\; .
 \label{twisttr}
 \end{eqnarray}

It is not hard to verify that in both cases the algebra of the
supersymmetry transformations closes {\it off shell}. As one can see from
(\ref{freetr}) and (\ref{twisttr}), the main difference between the two
types of supersymmetry amounts to interchanging
different types of $SU(2)$
indices (``twist"). Therefore we shall refer to the transformations
(\ref{freetr}) as {\it non-twisted} and to (\ref{twisttr}) as {\it twisted}
(4,4) supersymmetry. It is a well-known fact that dimensional reduction
from $N=2 D=4$ gives rise to the former, whereas the latter is specific to
two dimensions \cite{GHR}, \cite{IK}.

The main question now is whether we can turn on a background in the free
action (\ref{freeact}) compatible with any of the above supersymmetries.
The advantage of dealing with off-shell supersymmetry is that we do not have to
adjust the transformation laws to the interaction. As will be clear from
the end results, a simple self-dual Yang-Mills background like in (\ref{ADga})
cannot be compatible with (4,4) supersymmetry. One is forced to introduce an
additional ``curved" deformation of the free action (this point has already
been
made clear in \cite{CHS}). In order not to miss any possibility,
we shall examine the most general background for the action
(\ref{freeact}) allowed by the Lorentz and $U(1)$ properties and dimensions of
the
superfields $X,P,\Lambda$ (note that $X$ is dimensionless and $[P]=1,
[\Lambda] = 1/2$). So, we write down
\begin{equation}\label{genact}
S = \int d^2xdud^2\theta^+_+\; \left[i {\cal L}^{+Y}\partial_{++} X^+_Y +
iP^{-Y}_{++}
(D^{++} X^+_Y + {\cal L}^{+3}_Y)
+ {1\over 2} \Lambda^{i}_+ (\delta^{ij}D^{++} +
(V^{++})^{ij})\Lambda_{+}^{j} \right] \; .
\end{equation}
Here ${\cal L}^{+Y}(X^+,u)$, ${\cal L}^{+3Y}(X^+,u)$ and
$(V^{++})^{ij}(X^+,u)$ are for the time being arbitrary functions of
$X^{+Y}$ and $u^\pm_A$.

The next step is to vary the action (\ref{genact}) under either
(\ref{freetr}) or (\ref{twisttr}), derive the corresponding restrictions
on the potentials ${\cal L}^{+Y}$, ${\cal L}^{+3Y}$, $(V^{++})^{ij}$
and solve them in terms of unconstrained prepotentials. The computations are
straightforward, therefore here we shall only give the final answers. Up to
insignificant
field redefinitions those are:
\subsection {\it Non-twisted case}
Here the potential ${\cal L}^{+Y}$ takes the
form (after field redefinitions) ${\cal L}^{+Y} = X^{+Y}$. The other two
potentials
are expressed in terms of a single scalar prepotential with $U(1)$ charge $+4$
${\cal
L}^{+4}(X^+,u)$ as follows
\begin{equation}\label{potentials}
{\cal L}^{+3}_Y = \partial^-_Y {\cal L}^{+4}, \ \ \ (V^{++})_{\dot AY|\dot BZ}
=
-\epsilon_{\dot A\dot B} \partial^-_Y\partial^-_Z {\cal L}^{+4} \; .
\end{equation}
Once again, one realizes that this form of the action originates from the
dimensional reduction of the general $N=2\; D=4$ hypermultiplet action
\cite{HK}
\begin{equation}\label{hk}
S = \int d^4xdud^4\theta^+\; [q^{+Y}D^{++}q^+_Y + {\cal L}^{+4}(q^+,u)] \; .
\end{equation}
The prepotential ${\cal L}^{+4}$ in (\ref{hk}) has been shown in \cite{HK}
to generate
the most general hyper-K\"ahler manifolds. Such manifolds are torsion-free
and the connection term $(V^{++})_{\dot AY|\dot BZ}$ in (\ref{potentials}) is
in fact
the Christoffel connection.

\subsection {\it Twisted case} The requirement of (4,4) supersymmetry
of the type (\ref{twisttr}) leads to the following restrictions on the
potentials in (\ref{genact}):
\begin{equation}\label{twpot} {\cal L}^{+3}_Y = 0\; , \ \ \ (V^{++})_{A\dot
Y|B\dot Z} =
\epsilon_{\dot Y\dot Z} u^+_Au^+_B[1- V(X^+,u)]\; , \ \ \ \partial^-_Y {\cal
L}^{+}_Z -
\partial^-_Z {\cal L}^{+}_Y = -2 \epsilon_{YZ} V(X^+,u)\; .
\end{equation}
Once more we have a single scalar prepotential $V(X^+,u)$, but this time
it carries no $U(1)$ charge. Note that eq. (\ref{twpot}) determines the
potential ${\cal L}^{+Y}$ up to a gradient $\partial^-_Y{\cal L}^{++}$,
but one can easily see that this is a gauge invariance of the action
(\ref{genact}). Written down in terms of the restricted potentials
(\ref{twpot}), the action compatible with twisted supersymmetry takes the
form
\begin{eqnarray}
S &=& \int d^2xdud^2\theta^+_+\; \left(i {\cal L}^{+Y}(X^+,u)\partial_{++}
X^+_Y +
iP^{-Y}_{++} D^{++} X^+_Y\right. \nonumber \\
& &\ \ \ \ \  \left. -\Lambda_+^{-\dot Y} D^{++}\Lambda^+_{+\dot Y} -{1\over
2}
V(X^+,u)\Lambda_+^{+\dot Y}
\Lambda^+_{+\dot Y} \right) \; .\label{twact}
\end{eqnarray}
We see that in fact this action is very similar to the one obtained in section
2. The difference is that in (\ref{twact}) the gauge group is reduced to
$SU(2)$ (the gauge superfield $V^{++}$ being given in (\ref{twpot})) and
the kinetic term for $X^+$ is deformed by the potential ${\cal L}^{+Y}$.
Note that the presence of this potential does not affect in any way
the arguments leading to the ADHM-type interaction in section 2; there
we have never used the kinetic term for $X^+$.

Finally, we briefly mention the generalization of the above results to the
case of $4k$ target space dimensions. In fact, the ADHM sigma model of
Witten \cite{Witten} reviewed in section 2 can equally well accommodate
$R^{4k}$ as its target space, just replacing the $Sp(1)$ spinor index $Y$
of $X^{+Y}$ by an $Sp(k)$ index. For our study of $(4,4)$ supersymmetry it
is convenient to adapt the notation as follows. Instead of $X^{+Y}$ we
write $X^{+Y\alpha}$, where $Y$ is still an $Sp(1)$ spinor index and we
have added a new vector index $\alpha$ of $SO(k)$ (thus $Y\alpha$ forms a
decomposition of an $Sp(k)$ spinor index under $Sp(1)\times SO(k)$). The
same index $\alpha$ will be attached to the chiral fermions, e.g.,
$\Lambda^{A\dot Y\alpha}_+$. This amounts to an obvious modification of
the $(4,4)$ supersymmetry transformation rules (\ref{freetr}) and
(\ref{twisttr}) and of the general interacting action (\ref{genact}).  In
the non-twisted case we find once again a hyper-K\"ahler background, hence
no gauge field. In the twisted case we obtain the analog of (\ref{twact})
\begin{eqnarray}
S &=& \int d^2xdud^2\theta^+_+\; \left(i {\cal
L}^{+Y\alpha}(X^+,u)\partial_{++} X^{+\alpha}_{Y} +
i P^{-Y\alpha}_{++} D^{++} X^{+\alpha}_{Y}
\right. \nonumber \\ & &\ \ \ \ \  \left.-
\Lambda_+^{-\dot Y\alpha} D^{++}\Lambda^{+\alpha}_{+\dot Y} -{1\over 2}
V_{\alpha\beta}(X^+,u)\Lambda_+^{+\dot Y\alpha}
\Lambda^{+\beta}_{+\dot Y} \right) \; .\label{twaction}
\end{eqnarray}
Here the potentials ${\cal L}^{+Y\alpha}(X^+,u)$ and
$V_{\alpha\beta}(X^+,u)=V_{\beta\alpha}(X^+,u)$ are related by the constraint
\begin{equation}
\partial^-_{Y\alpha} {\cal L}^+_{Z\beta} - \partial^-_{Z\beta} {\cal
L}^+_{Y\alpha} =
-2\epsilon_{YZ} V_{\alpha\beta}\; .
\end{equation}
We see that the gauge connection $V^{++}$ from (\ref{genact}) takes the form
$(V^{++})_{A\dot Y\alpha|B\dot Z\beta} = \epsilon_{\dot Y\dot Z} u^+_Au^+_B
[\delta_{\alpha\beta} -V_{\alpha\beta}]$, so it corresponds to the gauge group
$Sp(k)$ instead of $SU(2)\sim
Sp(1)$ in the four-dimensional case.

\section{Components of the twisted sigma model}

In order to better understand the content of the twisted sigma model
(\ref{twact}) we shall present its component expansion (in the case of 4
space-time
dimensions only). This is made very easy by the fact that the Lagrange
multipliers
$P^{-Y}_{++}$ and
$\Lambda_+^{-\dot Y}$ in (\ref{twact}) force
both superfields $X^+_Y $ and $\Lambda_+^{+\dot Y}$ to depend trivially on the
harmonics, therefore they contain only finite numbers of fields:
\begin{eqnarray}
X^{+Y} &=& X^{AY}(x)u^+_A + i\theta_+^{+A'}\psi_{-A'}^Y(x) -i
(\theta^+_+)^2 \partial_{--} X^{AY}(x)u^-_A \; , \nonumber \\
\Lambda_+^{+\dot Y} &=& \lambda_+^{A\dot Y}(x)u^+_A +
\theta_+^{+A'}s_{A'}^{\dot Y}(x)
-i (\theta^+_+)^2 \partial_{--} \lambda_+^{A\dot Y}(x)u^-_A \; .
\end{eqnarray}
We then insert these expansions in the action and do the Grassmann
integral. The field $s$ is easily seen to be auxiliary, so we eliminate it and
in the end obtain the following sigma model for the fields
$X^{AY},\psi_{-A'}^Y,\lambda_+^{A\dot Y}$:
\begin{eqnarray}
S &=& \int d^2x\; [-(\epsilon_{AB} g(X) + b_{AB}(X)) \partial_{++}X^{AY}
\partial_{--}X^{B}_Y
+{i\over 2}g(X) \psi_-^{A'Y}i\partial_{++} \psi_{-A'Y}   \nonumber
\\  &+& {i\over 2}g(X) \lambda_+^{A\dot Y}i\partial_{--}
\lambda_{+A\dot Y} + {i\over 2}\psi_-^{A'Y} \psi_{-A'}^T \partial_{++}X^{CZ}
\Omega_{CZ|YT}(X)
 \nonumber\\ &+& {i\over 2}\lambda_+^{A\dot Y} \lambda_{+\dot
Y}^{B}\partial_{--}X^{CZ}
\Omega_{CZ|AB}(X)  +
\lambda_+^{A\dot Y} \lambda_{+\dot Y}^{B} \psi_-^{A'Z} \psi_{-A'}^T R_{ABZT}(X)
]\; .
\label{compact}
\end{eqnarray}
Here the sigma model metric is given by
\begin{equation}\label{metric}
g_{AY|BZ}(X) =   \epsilon_{AB}\epsilon_{YZ}\; g(X)\ \ \mbox{\rm with} \ \
g(X) = \int du\; V(X^+,u)\; ;
\end{equation}
the two-form (the torsion potential) is
\begin{equation} b_{AY|BZ}(X) = \epsilon_{YZ} b_{AB}= 2\int du\;
u^+_{(A}u^-_{B)} V(X^+,u)\; ;
\end{equation} the spin connections are
\begin{equation}\label{spinc}
 \Omega_{CZ|YT}(X) =  \epsilon_{Z(T}\partial_{CY)}g, \;\;
\Omega_{CZ|AB}(X) = - \epsilon_{C(B}\partial_{A)Z}g\;.
\end{equation}
 Finally,  the curvature $R$ in the four-fermion term is constructed in the
usual way.

The action (\ref{compact}) has two remarkable properties. Firstly, we see
that the geometric objects - metric and torsion (but not the two-form
itself) are expressed in terms of a {\it single real scalar function}
$g(X^{AY})$. In particular, this means that the sigma model metric is {\it
conformally flat}. Secondly, the function $g(X^{AY})$ satisfies Laplace's
equation. This is obvious from the definition (\ref{metric}):
\begin{equation}
\Box g(X) = \int du\; \partial^{-Y} \partial^+_Y V(X^+,u) = 0
\end{equation}
because of the holomorphic dependence of the potential $V(X^+,u)$ on
$X^{+Y}$. In fact, what we see here is an example of Penrose's transform
\cite{Penrose}, where the solutions to Laplace's equation are parametrized
by an unconstrained holomorphic function in twistor space (the r\^ole of
twistor variables here is played by the $SU(2)$ harmonics). The spin
connections in (\ref{spinc}) consist of two parts - of the Riemannian
connection and of the torsion. As explained in \cite{CHS}, one of the two spin
connections
can also be viewed as an $SU(2)$ gauge field. As a consequence of (4,4)
off-shell supersymmetry this gauge field has the typical form of a 't
Hooft instanton solution \cite{Hooft}. Remembering that we started from
the action (\ref{ADga}), in which the gauge field was by construction of the
instanton (ADHM) type, we see that indeed we deal with a sigma model in
which 't Hooft's instantons appear in a natural way.

\section{Conclusions}

In this paper we have studied the conditions under which the massless $(0,4)$
sigma
model involving chiral fermions coupled to an ADHM gauge field and obtained by
Witten's procedure can have a larger $(4,4)$ supersymmetry. The main assumption
we
have made is that the interacting theory should preserve the off-shell $(4,4)$
supersymmetry of the free theory. We have seen that starting from the
non-twisted
free supersymmetry we could only obtain a hyper-K\"ahler sigma model without
torsion and, consequently, without a self-dual gauge field. If we choose to
preserve
the other, twisted supersymmetry of the free theory, we obtain strong
restrictions on
the possible background: it must have a conformally flat metric and torsion
expressed
in terms of a single real scalar function $g(X)$. The latter satisfies
Laplace's equation
and thus gives rise to an $SU(2)$ instanton gauge field of the 't Hooft type.

We should point out that one could reach the same conclusions by applying the
general
results of \cite{GHR} to the special case of a four-dimensional target space.
In
\cite{GHR} the analysis of the conditions for $(4,4)$ supersymmetry in a sigma
model
is carried out in terms of $(2,2)$ superfields (chiral and twisted chiral).
This implies
choosing holomorphic coordinates in the target space. The sigma model
Lagrangian is
given by a K\"ahler potential.  Our scalar function $g(X)$ appears in
\cite{GHR} as a
second-order derivative of the K\"ahler potential. The main difference between
this
approach and ours (which makes use of $(0,4)$ superfields) is in the treatment
of  the
$SU(2)$ symmetry inherent to the problem we address here. Working in  a
holomorphic
basis inevitably leads to loosing manifest $SU(2)$ invariance.

Another study of $(4,4)$ twisted sigma models has recently been presented in
\cite{IvSut}, this time using a double-harmonic $(4,4)$ superspace formalism.
The
results obtained there agree with ours. We believe that the $(0,4)$ approach
may prove
more efficient in investigating the possible {\it linear} $(4,4)$ sigma models
with
potential terms.

We would also like to note that the
relevance of 't Hooft instantons in string-inspired sigma models has been shown
very
clearly in \cite{CHS}. There one deals with two distinct cases: first with
$(0,4)$ and
then with $(4,4)$ supersymmetry. In the context of $(0,4)$ supersymmetry the
coupling
to a 't Hooft instanton gauge field (as opposed to a general, ADHM type one)
appears as
an Ansatz, which fits nicely with string theory. Then, for reasons having to do
with the
quantum behaviour of the model, the authors of
\cite{CHS} want to extend the $(0,4)$ supersymmetry to full $(4,4)$ and realize
that
the self-dual gauge field must necessarily be identified with the spin
connection with
torsion. Thus, a $(4,4)$ model incorporating a self-dual gauge field cannot
have a flat
geometry. A point which is missing in \cite{CHS} is the observation that in the
context
of $(4,4)$ supersymmetry 't Hooft instantons are not an Ansatz any more, but
are the
only possibility. We also remark that some conclusions reached in \cite{CHS}
about
the general conditions on the background for $(0,4)$ supersymmetry have been
later on
corrected in \cite{BonVal}.

In conclusion we may say that the study of instantons in the context of $(4,4)$
sigma
models in this paper should be considered as a first step only. The more
interesting
question is whether there exists a {\it linear} $(4,4)$ sigma model which
automatically gives, in the infrared limit, the model with 't Hooft instantons
discussed
here. An encouraging sign is the fact that the free action of section 2,
in which we only keep the mass term in (\ref{int}) (but drop the Yukawa
couplings), does
indeed have a $(4,4)$ supersymmetry. However, the massive $(4,4)$ multiplet has
central charge and is on shell, which makes the analysis of the general
potential
self-interaction more difficult. We hope to come back to this problem in the
near
future.

\vskip7mm
{\bf Acknowledgements.} We are grateful to E. Witten for encouraging us to look
into this problem. We also profited from discussions with C. Callan, J. Harvey,
C. Hull and M. Rocek. E. S. would like to acknowledge the hospitality extended
to him at the Johns Hopkins University, Baltimore and ITP, SUNY at
Stony Brook where this work has been done.

\end{document}